\documentclass{llncs}
\usepackage[english]{babel}
\usepackage[T1]{fontenc}
\usepackage{amsmath}
\usepackage{amssymb,amsmath}
\usepackage{graphicx}
\usepackage{caption}
\usepackage{wrapfig}
\usepackage{multicol}
\usepackage{calligra}
\usepackage{multirow}
\usepackage{url}

\begin{document}
\pagenumbering{arabic}
\title{Disease Progression Modeling and Prediction through Random Effect Gaussian Processes and Time Transformation }
\author{Marco Lorenzi \inst{1,2} \and Maurizio Filippone \inst{3} \and Daniel C. Alexander \inst{2} \and Sebastien Ourselin \inst{2} for  the Alzheimer's Disease Neuroimaging Initiative\thanks{Data used in preparation of this article were obtained from the Alzheimer's Disease Neuroimaging Initiative (ADNI) database (www.loni.usc.edu/ADNI). As such, the investigators within the ADNI contributed to the design and implementation of ADNI and/or provided data but did not participate in analysis or writing of this report. A complete listing of ADNI investigators can be found at: $\mbox{www.loni.usc.edu/ADNI/Collaboration/ADNI\_Authorship\_list.pdf}$}}
\institute{Asclepios Research Project, Universit\'e C\^ote d'Azur, Inria, France \and Centre for Medical Image Computing, University College London, UK  \and Department of Data Science, EURECOM, France}%Translational Imaging Group, CMIC, UCL, London, UK }
\maketitle

\begin{abstract} 
The development of statistical approaches for the joint modelling of the temporal changes of imaging, biochemical, and clinical biomarkers  is  of paramount importance for improving the understanding of neurodegenerative disorders, and for providing a reference for the prediction and quantification of the pathology in unseen individuals. Nonetheless, the use of disease progression models for probabilistic predictions still requires investigation, for example for accounting for missing observations in clinical data, and for accurate uncertainty quantification. 
We tackle this problem by proposing a novel Gaussian process-based method for the joint modeling of imaging and clinical biomarker progressions from time series of individual observations. The model is formulated to account for individual random effects and time reparameterization, allowing non-parametric estimates of the biomarker evolution, as well as high flexibility in specifying correlation structure, and time transformation models. Thanks to the Bayesian formulation, the model naturally accounts for missing data, and allows for uncertainty quantification in the estimate of evolutions, as well as for probabilistic prediction of disease staging in unseen patients. 
%The model is benchmarked on synthetic data and validated on multimodal imaging and clinical measurements from a cohort of 517 amyloid positive individuals of the ADNI dataset. 
The experimental results show that the proposed model 
%can reliably estimate biomarker temporal progressions as well as individual time parameters under different testing conditions. It also 
provides a biologically plausible description of the evolution of Alzheimer's pathology across the whole disease time-span as well as remarkable predictive performance when tested on a large clinical cohort with missing observations. %, thus representing a promising instrument for the analysis of clinical trials data beyond the state-of-art.
\end{abstract} 

\section{Introduction}

Neurodegenerative disorders (NDDs), such as Alzheimer's disease (AD), are characterised by the progressive pathological alteration of the brain's biochemical processes and morphology, and ultimately lead to the irreversible impairment of cognitive functions.
%Since the aetiology of NDDs is represented by complex and multifactorial biological alterations, 
The correct understanding of the relationship between the different pathological features is of paramount importance for improving the identification of pathological changes in patients, and for better treatment. To this end, the recent availability of collections of imaging and clinical data in NDDs is a unique opportunity to define statistical models for  the {joint modelling} of the temporal changes of {imaging}, {biochemical}, and {clinical} biomarkers.

The goal of disease progression modeling in NDDs is twofolds: 1) \emph{quantifying} the dynamics of NDDs along with the related temporal relationship between different biomarkers, and 2) \emph{staging} patients based on individual observations for diagnostic and interventional purposes. 
The related challenge is in the definition of optimal methods to integrate and jointly analyze the heterogeneous and highly multi-modal information available to clinicians. Moreover, longitudinal clinical datasets of NDDs generally lack of a well-defined temporal reference, since the onset of the pathology may vary across individuals according to genetic and environmental factors \cite{yang2011}. Therefore, age or visit date information are biased time references for the individual longitudinal measurements. 

To tackle this problem, in \cite{donohue2014} the authors proposed to model the temporal biomarker trajectories as a random effect regression problem, building on the well established theory of {self-modeling regression} \cite{kneip1988}. Practically, the trajectories are modelled by monotonic B-splines, and the estimation is performed by subsequent minimization of the partial residuals sum of squares associated with regression parameters and individual time shift, respectively. %Since the biomarkers progressions are modelled through monotonic splines interpolation of the random effects, this approach does not allow to consistently model the trajectory uncertainties. 
Based on the assumption of a logistic curve shape for the average biomarker trajectories, \cite{schiratti2015} frames the random effect regression model in a Riemannian setting, in which the random effects identify individual time-shift and acceleration factors. 
Finally, image-based progression models have been recently proposed \cite{younes2014,bilgel2015}, based on time-reparameterization of voxel/mesh-based measures. 
While the main focus of current works mainly concerns the modeling of neurodegeneration, the use of disease progression models for predictive purposes is much less investigated. Predictive models of patient staging were proposed within the setting of event based models \cite{fonteijn2011}, or still through random effect modeling \cite{guerrero2016}. However, the event based model relies on the coarse binary discretization of the biomarker changes, and does not account for longitudinal observations, while the model proposed in \cite{guerrero2016} requires cohorts with known disease onset, and therefore does not easily generalise to the study of general clinical populations. 

In this study we propose an unified approach to the consistent disease progression modeling and probabilistic prediction of clinical data, by introducing a Bayesian regression framework 
%To this purpose we propose a novel probabilistic approach for the joint modeling 
of imaging and clinical biomarker progressions from time series of individual observations.
The model is based on Gaussian process (GP) regression, and is formulated to account for individual random effects and time reparameterization, as well to naturally account for missing observations. The {methodological contribution} of this study consists in reformulating disease progression modeling within a Bayesian context, allowing non-parametric estimates of the biomarker evolution, as well as high flexibility in specifying random effects structure, and time reparameterization models. The uniqueness of the time transformation is enforced by imposing a {monotonicity constraint} on the trajectories, via a prior on the temporal derivatives, while  the model is computationally efficient thanks to the block-wise algebraic structure of the GP covariance function. 
From the application point of view, the proposed approach enables novel applications of disease progression modeling such as the \emph{probabilistic prediction} of disease staging in unseen patients. Moreover, the model naturally accounts for missing data, and allows for uncertainty quantification of both evolutions and parameters.

Section \ref{section:Model} introduces the proposed GP regression model of joint temporal progression of biomarkers, along with the prior derivatives specification.  The resulting posterior and approximated inference scheme through expectation propagation (EP) is described in Section \ref{section:Likelihood}. Finally, Section \ref{section:experiments} illustrates the validation of our model on i) synthetic data and ii) clinical and multivariate imaging measurements from a cohort of 517 amyloid positive individuals of the ADNI database. 
%The synthetic benchmarking shows that the proposed model can reliably estimate biomarker temporal progression under different testing conditions.  Finally, when applied to the ADNI data 
The experimental results show that our model provides a biologically plausible description of the evolution of AD across the whole disease time-span, as well as statistically significant prediction of disease staging and high classification accuracy on unseen and incomplete testing data.

\section{Gaussian process-based random effect modeling of  longitudinal progressions}\label{section:Model}
In what follows, longitudinal measurements of $N_b$ biomarkers $\{b_1, \ldots, b_{N_b}\}$ over time are given for $N$ individuals. %, and we assume that the number of measurements may vary across individuals and biomarkers.  

We represent the longitudinal biomarker's measures associated with each individual $j$ as a multidimensional array $(\mathbf{y}^j({t_1}), \mathbf{y}^j({t_2}), \ldots, \mathbf{y}^j({t_{k^j}}))^{\top}$ sampled at $k^j$ multiple time points $ \mathbf{t} = \{{t}_{1}, {t}_{2}, \ldots, {t}_{k^j}\}$. Although different biomarkers may be in reality sampled at different time-points, for sake of notation simplicity  in what follows we will assume, without loss of generality, that the sampling time is common among them. The observations for individual $j$ at a single time point $t$ are thus a random sample from the following generative model:
\begin{align}\label{GenerativeModel}
{\mathbf{y}^j(t)} = \left({{y}_{b_1}^j(t)}, {{y}_{b_2}^j(t)}, \ldots, {y}_{b_{N_b}}^j(t)\right)^{\top} &= \mathbf{f}({t}) + {\boldsymbol{\nu}^j}({t}) + \boldsymbol{\epsilon},
\end{align}
where $\mathbf{f}({t}) = ({{f}_{b_1}}({t}), {{f}_{b_2}}({t})\ldots, {{f}_{b_{N_b}}}({t}))^{\top}$ is the fixed effect function modelling the biomarker's longitudinal evolution, $\boldsymbol{\nu}^j({t}) = ({\nu}_{b_1}^j({t}),{\nu}_{b_2}^j({t}), \ldots, {\nu}_{b_{N_b}}^j({t}))^{\top}$ is the individual random effect, and $\boldsymbol{\epsilon} = ({\epsilon}_{b_1},{\epsilon}_{b_2}, \ldots,{\epsilon}_{b_{N_b}})^{\top} $ is observational noise independent from time.   
The group-wise evolution is modelled as a zero-mean GP, $\mathbf{f}\sim \mathcal{GP}(0,\Sigma_G)$, the individual random effects are assumed to be Gaussian distributed correlated signal $\boldsymbol{\nu}^j\sim \mathcal{N}(0,\Sigma_S)$, while the observational noise is assumed to be a Gaussian heteroskedastic term $\boldsymbol{\epsilon} \sim \mathcal{N}(0,\Sigma_{\epsilon})$, where $\Sigma_{\epsilon}$ is a diagonal matrix  $\mbox{diag}[\boldsymbol{\sigma}_{b1}^2,\boldsymbol{\sigma}_{b2}^2, \ldots, \boldsymbol{\sigma}_{b_{N_b}}^2 ]$.\\
\textbf{Fixed Effect Process.} The covariance function $\Sigma_G$ describes the biomarkers temporal variability, and is represented as a block-diagonal matrix  
\begin{eqnarray}\label{FixCovariance}
\Sigma_G(\mathbf{f},\mathbf{f}) &= \mathrm{diag}[ \Sigma_{b_1}(\mathbf{f}_{b_1},\mathbf{f}_{b_1}), \Sigma_{b_2}(\mathbf{f}_{b_2},\mathbf{f}_{b_2}), \ldots, \Sigma_{b_{N_b}}(\mathbf{f}_{b_{N_b}},\mathbf{f}_{b_{N_b}}) ],
\end{eqnarray}
where each block represents the within-biomarker temporal covariance expressed as a negative squared exponential function:
\begin{align}
\Sigma_{b}(\mathbf{f}_b(t_1),\mathbf{f}_b(t_2)) &= \eta_b \exp\left(-\frac{(t_1-t_2)^2}{2\,l_b^2}\right),    
\end{align}
and where the parameters $\eta_b$ and $l_b$ are the marginal variance and length-scale of the biomarker's temporal evolution, respectively. \\
\textbf{Individual Random Effects.} 
The random covariance function $\Sigma_S$ models the individual deviation from the fixed effect, and is represented as a block-diagonal matrix 
$$\Sigma_S (\boldsymbol{\nu}^j,\boldsymbol{\nu}^j)= \mathrm{diag}[\,\Sigma_{b_1}^j(\boldsymbol{\nu}_{b_{1}}^j,\boldsymbol{\nu}_{b_{1}}^j), \Sigma_{b_2}^j(\boldsymbol{\nu}_{b_{2}}^j,\boldsymbol{\nu}_{b_{2}}^j), \ldots, \Sigma_{b_{N_b}^j}(\boldsymbol{\nu}_{b_{N_b}}^j,\boldsymbol{\nu}_{b_{N_b}}^j) ],$$
where each block $\Sigma_{b}^j$ corresponds to the covariance function associated with the individual process $\boldsymbol{\nu}_{b}^j({t})$. Thanks to the flexibility of the proposed generative model, any form of the random effect covariance $\Sigma_S$ can be easily specified in order to model the subject-specific biomarkers' progression. In what follows we will use a linear covariance form $\Sigma_{b}^j(\boldsymbol{\nu}_{b}^j(t_1), \boldsymbol{\nu}_{b}^j(t_2)) = (\sigma_b^j)^2 \left((t_1 - \overline{\mathbf{t}}) (t_2 - \overline{\mathbf{t}}) \right)$, where $\overline{\mathbf{t}}$ is the average observational time for individual $j$, if more than 4 measurements were available, and i.i.d. Gaussian covariance form $\Sigma_{b}^j(\boldsymbol{\nu}_{b}^j(t_1), \boldsymbol{\nu}_{b}^j(t_2)) = (\sigma_b^j)^{2}$ if 2 or 3 measurements were available, while assigning it to 0 otherwise. This choice is motivated by stability concerns, in order to keep the model complexity compatible with the generally limited number of measurements available for each individual. \\
\textbf{Individual time transformation.}
The generative model (\ref{GenerativeModel}) is based on the key assumption that the longitudinal observations across different individuals are defined with respect to the same temporal reference. This assumption may be invalid when the temporal alignment of the individual observations with respect to the common group-wise model is unknown, for instance in the typical scenario of a clinical trial in AD where the patients' observational time is relative to the common baseline, and where the disease onset is a latent event (past or future) which is not directly measurable.
We assume that each individual measurement is made with respect to an absolute time-frame $\tau$ through a time-warping function $t = \phi^j(\tau)$ that models the time-reparameterization with respect to the common group-wise evolution. Model (\ref{GenerativeModel}) can thus be reparameterized as 
\begin{align}\label{GenerativeModelNew}
\mathbf{y}^j(\phi^j(\tau)) = \mathbf{f}(\phi^j(\tau)) + \boldsymbol{\nu}^j(\phi^j(\tau)) + \boldsymbol{\epsilon}. 
\end{align}

The present formulation allows the specification of any kind of time transformation, and in what follows we shall focus on the modelling of a linear reparameterization of the observational time  $\phi^j(\tau) = \tau + d^j$. This modeling assumption is mostly motivated by the choice of working with a reasonably limited number of parameters, compatibly with the generally short follow-up time available per individual (cfr. Table \ref{Nobs}). Within this setting, the time-shift $d^j$ encodes the disease stage associated with the individual relatively to the group-wise model. 

Overall, model (\ref{GenerativeModelNew}) is  identified by $(N_j + 3) N_b + N_j $ parameters, represented by the fixed effects and noise $\boldsymbol{\theta}_G = \{\eta_{b_k}, l_{b_k}, \epsilon_{b_k}\}_{k=1}^{N_b}$, by the individual random effects parameters $\boldsymbol{\theta}_G^j = \{\sigma_{b_k}^j\}_{k=1}^{N_b}$ and by the time-shifts $d^j$. \\
\textbf{Monotonic constraint in multimodal GP regression.} 
Due to the non-parametric nature of Gaussian process regression, we need an additional constraint on model (\ref{GenerativeModelNew}) in order to identify a unique solution for the time-warp. By assuming  a steady temporal evolution of biomarkers from normal to pathological values, we shall assume that the biomarker trajectories described by (\ref{GenerativeModelNew}) follow a (quasi) monotonic behaviour. This requirement can be implemented by imposing a prior positivity constraint on the derivatives of the GP function $f$.  
Inspired by \cite{riihimaki2010gaussian}, we impose a monotonicity constraint by assuming a probit-likelihood for the  derivative measurements $\mathbf{m}(t)$ associated with the derivative process $\mathbf{\dot{f}}(t) = \frac{\mathrm{d} \mathbf{f}(t)}{\mathrm{d} t}$ at time $t$:
\begin{align}
p(\mathbf{m}(t)|\mathbf{\dot{f}}(t)) &= \Phi\left(\frac{1}{\lambda}\mathbf{\dot{f}}(t)\right),
\end{align}
with $\Phi(z) = \int_{-\infty}^z\mathcal{N}(x|0,1)\,dx$. The quantity $\lambda > 0$ is an additional model parameter controlling the degree of positivity enforced on the derivative process, with values approaching to zero for stronger monotonicity constraint. In what follows, the monotonicity of each biomarker is controlled by placing 10 derivative points equally spaced on the observation domain, and by fixing the $N_b$ derivative parameters $\{\lambda_{b_k}\}_{k=1}^{N_b}$ to the value of $e$-6. This choice for the parameter is motivated by the need to enforce a strong monotonicity constraint, necessary for the stability of the initial time-shift estimation. \\
\textbf{On adding a monotonic constraint to the individual observations.} By following a similar construction, we could equally enforce a monotonic behaviour to the random effects associated with the individual trajectories. This additional constraint would however come with a cumbersome increase of the model complexity, since it would introduce an additional layer of virtual derivative parameters (with associated location) per individual. Moreover, while we are interested in modeling a {globally} monotonic biomarker trajectory on the fixed parameters, we relax this constraint at the individual level, since some subjects may be characterised by non strictly monotonic time-series due to specific clinical conditions.  

%\subsubsection{Model parameters.} Model (\ref{GenerativeModelNew}) is  identified by $(N_j + 3) N_b + N_j $ parameters, represented by the fixed effects and noise $\boldsymbol{\theta}_G = \{\eta_{b_k}, l_{b_k}, \epsilon_{b_k}\}_{k=1}^{N_b}$, by the individual random effects parameters $\boldsymbol{\theta}_G^j = \{\sigma_{b_k}^j\}_{k=1}^{N_b}$ and by the time-shifts $d^j$. 

\section{Joint Model: marginal likelihood and inference}\label{section:Likelihood}
Given the sets of individual biomarker measurements $\mathbf{y}=\{(\mathbf{y}^j(t_i))_{i=1}^{k^j}\}_{j=1}^{N}$, and of $D$ control derivatives $\mathbf{m}=\{m_{b_k}(t_l')\}_{l=1}^D$ at points $t' = \{t_l'\}_{l=1}^D$ for the progression of each biomarker $b_k$, the random effect GP model posterior is:      
\begin{align}
p\left(\mathbf{f},\mathbf{\dot{f}},\boldsymbol{\nu}^j|\mathbf{y},\mathbf{m} \right) &= \frac{1}{Z}p(\mathbf{f},\mathbf{\dot{f}}|t,t')p(\boldsymbol{\nu}|t)p(\mathbf{y}|\mathbf{f},\boldsymbol{\nu})p(\mathbf{m}|\mathbf{\dot{f}})\nonumber\\
                   &= p(\mathbf{f},\mathbf{\dot{f}}|t,t')p(\boldsymbol{\nu}|t)p(\mathbf{y}|\mathbf{f},\boldsymbol{\nu})\prod_k\prod_l\Phi\left(\frac{1}{\lambda}{\dot{f}}_{b_k}(t_l')\right)\label{posterior},
\end{align}
where $\boldsymbol{\nu} = \{\mathbf{\nu}^j\}_{j=1}^N$. Thanks to the linearity of GPs under derivation, we have that $Cov\left(\mathbf{f}(t),\mathbf{\dot{f}}(t')\right) = \frac{\mathrm{d}Cov(\mathbf{f}(t),\mathbf{f}(t'))}{\mathrm{d}t'}$, and that the joint distribution $p\left(\mathbf{f},\mathbf{\dot{f}}|t,t'\right)$ is again a GP:

\begin{eqnarray}\label{joint}
p\left(\mathbf{f},\mathbf{\dot{f}},\boldsymbol{\nu}^j|t,t'\right) &\sim& \mathcal{GP}\left(\mathbf{f}_{joint}|0,\Sigma_{joint}\right) \nonumber\\
\mathbf{f}_{joint}  = \begin{pmatrix}\mathbf{f}\\
\mathbf{\dot{f}}
\end{pmatrix} & \sim & \mathcal{N}\left[\left(\begin{array}{c}
0\\
0
\end{array}\right), \left(\begin{array}{cc}
\Sigma_G(\mathbf{f}(t),\mathbf{f}(t)) & \frac{\partial\Sigma_G(\mathbf{f}(t),\mathbf{f}(t'))}{\partial t'}\\
\frac{\mathrm{d}\Sigma_G(\mathbf{f}(t'),\mathbf{f}(t))}{\mathrm{d} t'} & \frac{\mathrm{d}^2\Sigma_G(\mathbf{f}(t'),\mathbf{f}(t'))}{\mathrm{d}{t'}^2}
\end{array}\right)\right]\enspace.\nonumber
\end{eqnarray}
\subsection{Approximated inference with Expectation Propagation}
Due to the non-Gaussianity of the derivative likelihood term, the direct inference on the posterior (\ref{posterior}) is not possible due to its analytically intractable form. For this reason, we employ an approximate inference scheme based on expectation propagation (EP) \cite{rasmussen2006gaussian,minka2001expectation}. Following \cite{riihimaki2010gaussian}, we compute an approximated posterior distribution $q\left(\mathbf{f},\mathbf{\dot{f}},\boldsymbol{\nu}^j|\mathbf{y}^j,\mathbf{m}\right)$ by replacing the derivative likelihood terms with local un-normalized Gaussian approximations:
\begin{align}
q\left(\mathbf{f},\mathbf{\dot{f}},\boldsymbol{\nu}^j|\mathbf{y}^j,\mathbf{m}\right) &=\frac{1}{Z_{EP}} p(\mathbf{f},\mathbf{\dot{f}}|t,t')p(\boldsymbol{\nu}|t)p(\mathbf{y}|\mathbf{f},\boldsymbol{\nu})\prod_k\prod_l \tilde{Z}_{kl}\mathcal{N}(\dot{f}_{b_k}(t_l')|\tilde{\mu}_{kl},\tilde{\sigma}^2_{kl}),
\end{align}
where $\prod_k\prod_l \tilde{Z}_{kl}\mathcal{N}(\dot{f}_{b_k}(t_l')|\tilde{\mu}_{kl},\tilde{\sigma}^2_{kl}) = \mathcal{N}(\boldsymbol{\tilde{\mu}},\tilde{\Sigma})\prod_{k,l}\tilde{Z}_{kl}$, with $\boldsymbol{\tilde{\mu}} = [\tilde{\mu}_{kl}]$, and $\tilde{\Sigma}$ is a  diagonal matrix with elements $\tilde{\sigma}^2_{kl}$. It follows that the marginal posterior has a Gaussian form, $q\left(\mathbf{f},\mathbf{\dot{f}},\boldsymbol{\nu}^j|\mathbf{y}^j,\mathbf{m}\right)\sim\mathcal{N}(\boldsymbol{\mu},\Sigma)$, with $\boldsymbol{\mu} = \Sigma\tilde{\Sigma}^{-1}\boldsymbol{\tilde{\mu}}_{joint}$ , and $\Sigma = (\Sigma_{joint}^{-1} + \tilde{\Sigma}_{joint}^{-1})^{-1}$, where 
\begin{eqnarray}
\boldsymbol{\tilde{\mu}}_{joint}  = \begin{pmatrix}\mathbf{y}\\
\boldsymbol{\tilde{\mu}}
\end{pmatrix} &\mbox{, and }  & \tilde{\Sigma}_{joint} = \left(\begin{array}{cc}
\Sigma_{\epsilon} + \Sigma_{S} & 0\\
0 & \tilde{\Sigma}
\end{array}\right)\enspace.
\end{eqnarray}

\subsubsection{Estimating the EP parameters.} The EP update of the local Gaussian approximation parameters is classically done by iterative moment matching with respect to the product between the cavity distributions $q_{-k'l'}\left(\dot{f}_{b_{k'}}(t'_{l'})\right)$ and the target likelihood term $\Phi\left(\frac{1}{\lambda}\dot{f}_{b_{k'}}(t'_{l'})\right)$.
In the GP case the cavity distribution has a straightforward Gaussian form:
\begin{eqnarray}
q_{-k'l'}\left(\dot{f}_{b_{k'}}(t'_{l'})\right) &=& \int \prod_{k\neq k'}\prod_{l\neq l'} \tilde{Z}_{kl}\mathcal{N}(\dot{f}_{b_k}(t_l')|\tilde{\mu}_{kl},\tilde{\sigma}^2_{kl}) d\dot{f}_{b_k}(t_l')\nonumber\\
&\sim&\mathcal{N}(\dot{f}_{b_{k'}}(t_{l'}')|\mu_{-k'l'},\sigma_{-k'l'}).\label{cavity}
\end{eqnarray}
As shown in \cite{riihimaki2010gaussian} for univariate monotonic regression, moments and updates of the approximation parameters can be computed in an analogous manner as in the classical GP classification problem \cite{rasmussen2006gaussian}.

\subsection{Marginal Likelihood and hyper-parameter estimation}
The model's log-marginal likelihood under the EP approximation is:
\begin{eqnarray}\label{final_posterior}
\log\mathcal{L} &=& -\frac{1}{2}\log|\Sigma_{joint} + \tilde{\Sigma}_{joint}| - \frac{1}{2}\boldsymbol{\tilde{\mu}}_{joint}^T(\Sigma_{joint} + \tilde{\Sigma}_{joint})^{-1}\boldsymbol{\tilde{\mu}}_{joint} +\nonumber\\ &&\sum_{k}\sum_{l}\frac{(\mu_{-kl}-\tilde{\mu}_{kl})^2}{2(\sigma_{-kl}^2)+\tilde{\sigma}_{kl}^2)} + \sum_k\sum_l \log\Phi(\frac{\mu_{-kl}}{\sqrt{\lambda_{k}^2+\sigma_{-kl}^2)}})+\nonumber\\
&&\frac{1}{2}\sum_k\sum_l\log(\sigma_{-kl}^2+\tilde{\sigma}_{kl}^2).
\end{eqnarray}
In what follows, the optimal parameters are obtained by maximising $\log\mathcal{L}$ through conjugate gradient descent, via  alternate optimization between the hyper-parameters $\boldsymbol{\theta}_G$ and $\boldsymbol{\theta}_G^j$, and the individuals' time-shifts $d^j$. The position of the derivative points was updated at each iteration, according to the changes of the GP domain. Regularisation was also enforced by introducing Gaussian priors for the parameters $\boldsymbol{\theta}_G$  and $\boldsymbol{\theta}_G^j$. We note that the block structure of the GP covariance allows the computation of the gradients with respect to the biomarkers' and individual parameters by working on matrices of much smaller dimension than the one of the whole GP, thus considerably improving the numerical stability and the computational efficiency of the optimization procedure. 

\subsection{Prediction of observations and individual staging}
Gaussian processes naturally allow probabilistic predictions given the observed data. At any given time point $t^*$, the posterior biomarker distribution has the Gaussian form $p(\mathbf{f}^*|t^*,\mathbf{y}, t, \mathbf{m}, t') \sim \mathcal{N}(\mathbf{f}^*|\boldsymbol{\mu}^*, \Sigma^*)$ with parameters:

\begin{eqnarray}
\boldsymbol{\mu^*} &=& \Sigma_G(\mathbf{f}(t^*),\mathbf{f}(t))(\Sigma_{joint} + \tilde{\Sigma}_{joint})^{-1}\boldsymbol{\tilde{\mu}}_{joint}\\
\Sigma^* &=& \Sigma_G(\mathbf{f}(t^*),\mathbf{f}(t^*)) - \Sigma_G(\mathbf{f}(t^*),\mathbf{f}(t))(\Sigma_{joint} + \tilde{\Sigma}_{joint})^{-1}\Sigma_G(\mathbf{f}(t),\mathbf{f}(t^*)).
\end{eqnarray}
We also derive a probabilistic model for the individual temporal staging  given a set of biomarker observations $\mathbf{y}^*$, thanks to the Bayes formula:
\begin{eqnarray}
p(t^*|\mathbf{y}^*,\mathbf{y}, t, \mathbf{m}, t') = p(\mathbf{y}^*|t^*,\mathbf{y}, t, \mathbf{m}, t')p(t^*)/p(\mathbf{y}^*|\mathbf{y}, t, \mathbf{m}, t'), 
\end{eqnarray}
which we compute by assuming an uniform distribution on $t^*$, and by noting that $p(\mathbf{y}^*|t^*,\mathbf{y}, t, \mathbf{m}, t')~\sim\mathcal{N}(\boldsymbol{\mu}^*,\Sigma^*+\Sigma_{\epsilon})$. 
In particular, the joint covariance form $\Sigma_G(\mathbf{f}(t^*),\mathbf{f}(t^*))$ can be specified in order to account for incomplete data, and thus generalizes the GP model for predictions in presence of \emph{missing biomarker observations}. 

\section{Experiments}\label{section:experiments}
\subsection{Synthetic multivariate progressions}
We benchmarked the model with respect to synthetic multivariate biomarker progressions.
We generated random multivariate sigmoid functions for $N_b$ biomarkers, $\mathbf{f}(\tau) = (f_{b_1}(\tau), f_{b_2}(\tau), \ldots, f_{b_{N_b}}(\tau))^{\top}$, with $f_{b_k}(\tau) = 1/(1+\exp(-\alpha_k \tau))$, $\tau\in [0,15]$ and $\alpha_k \sim \mathcal{N}(0,.06) $, and we sampled $N$ individual noisy trajectories at time points $\tau_k^j$: $\mathbf{y}_k^j(\tau_k^j) = f_k(\tau_k^j) + \epsilon$, $\epsilon\sim\mathcal{N}(0,\sigma^2)$.  For each individual we used the same initial sampling time point for every biomarker, while the number of samples per biomarker was allowed to independently vary between 1 and 4. The individual time points were subsequently centered by their mean $\mu_k^j$ to obtain shifted time-points $t_k^j = \tau_k^j - \mu_k^j$ defined in the interval $[-2,2]$. 

The model was applied to estimate biomarker progressions and individual time-shifts with respect to different combinations of trajectory noise $\sigma$, sample size $N$, and number of biomarkers $N_b$. The accuracy of the model in reconstructing the original time series was quantified by Pearson's correlation between the estimated time-shift $d^j$ and the original individual time reference. The experiments were repeated 10 times for each configuration of parameters $\sigma \in \{0.1, 0.2, 0.3, 0.4\}$, $N_b\in\{4, 8\}$, and $N\in\{20,100\}$.\\
\begin{table}[b]
\centering
\begin{tabular}{c c c c c c || c  c c c c }
&&\multicolumn{4}{c}{$N = 20$}&\multicolumn{4}{c}{$N = 100$}\\\cline{3-10}
&&\multicolumn{4}{|c||}{$\sigma $}&\multicolumn{4}{c|}{$\sigma $} \\
 & \multicolumn{1}{c|}{}& 0.1 & 0.2 & 0.3 & 0.4 & 0.1 & 0.2 & 0.3 & \multicolumn{1}{c|}{.4} \\\cline{2-10}
\multirow{2}{*}{$N_b$} & \multicolumn{1}{c|}{4\,} &\multicolumn{1}{c|}{ .95 (.03)}  &\multicolumn{1}{c|}{ .86 (.08)} &\multicolumn{1}{c|}{ .71 (.17)} &\multicolumn{1}{c||}{ .46 (.29) } & \multicolumn{1}{c|}{ .91 (.04)} & \multicolumn{1}{c|}{.89(.04)} & \multicolumn{1}{c|}{.76 (.17)} & \multicolumn{1}{c|}{.75 (.12)}\\
& \multicolumn{1}{c|}{8\,} &\multicolumn{1}{c|}{ .97 (.01)} & \multicolumn{1}{c|}{.91 (.06) }& \multicolumn{1}{c|}{.86 (.06)} & \multicolumn{1}{c||}{.66 (.3)} & \multicolumn{1}{c|}{.94 (.04)} & \multicolumn{1}{c|}{.94 (.02)} & \multicolumn{1}{c|}{.88 (.06)} & \multicolumn{1}{c|}{.84 (.07)}\\\cline{2-10}
\end{tabular}\caption{Mean (sd) $R^2$ correlation coefficient across folds between estimated individual time-shifts and ground truth time reference.}\label{TableSynth}
\end{table}
{\bf Results.} Table \ref{TableSynth} reports summary correlations between time-shift estimation and the ground truth individual sampling time. The correlation values are generally high, and increase with lower noise levels. Interestingly, the increase in number of modelled biomarkers is associated with a better performance in recovering the underlying disease staging. We also observe that larger sample sizes are associated with higher correlation values, especially with increasing noise levels. We note however an exception for the case $\sigma = 0.1$ where, although the overall performance is still high, the correlation slightly decreases with $N=100$.

\subsection{Longitudinal modelling of Alzheimer's disease progression}
We collected longitudinal measurements for the ADNI individuals with baseline values of CSF A$\beta$ amyloid lower than the nominal values of 192 pg/ml. This preliminary selection is aimed to validate the model on a clinical population likely to represent the whole disease time-span.

The model was trained on a group including normal individuals, mild cognitive impairment subjects converted to AD (MCI conv), and AD patients having at least one measurement for each of the following biomarkers: \emph{volumetric measures} (hippocampal, ventricular, entorhinal, and whole brain volumes), \emph{glucose metabolism} (average normalized FDG uptake in prefrontal cortex, anterior cingulate, precuneus and 
parietal cortex),  \emph{brain amyloidosys} (average normalized AV45 uptake in frontal cortex, anterior cingulate, precuneus and 
parietal cortex), and \emph{cognitive function} measured by common cognitive questionnaires (ADAS13, RAVLT learning score, FAQ). The testing set was composed by the remaining subjects with at least a missing biomarker, as well as by the subgroup of MCI non converted to AD during the observational time (MCI stable).
The volumetric measures were scaled by the individual total intracranial volume, and all the biomarkers measurements were converted into quantile scores (0 to 1 for normal to abnormal values). 
Table \ref{sociodem} shows baseline clinical and sociodemographic information of the individuals used respectively in training and testing set, while in Table \ref{Nobs} we report the average follow-up time and the ratio of missing data of the pooled sample.

\begin{table}[t]
\centering
\begin{tabular}{ c c c c c c c}
Group & N & Age & Sex (\% females) & ADAS13 & Hippo volume ($mm^3$)& AV45  \\
\hline
\multicolumn{7}{c}{Training data}\\
\hline
NL & 76 & 75.8 (6) & 53 & 9.5 (4.4) & 7358 (762) & 1.24 (0.2)\\
MCI conv& 57 & 72.7 (7) & 42 & 20.3 (6.8) & 6464 (861) & 1.44 (0.2)\\
AD & 21 & 72.7 (10) & 43 & 29.3 (8.7) & 5872 (988) & 1.4 (0.2)\\
\hline
\multicolumn{7}{c}{Testing data}\\
\hline
NL & 30 & 77.5 (6) & 43 & 11.1 (4.1) & 7137 (800) & 1.03 (0.14)\\
MCI stable& 164 & 73.4 (6.9) & 39 & 15.5 (6.19) & 7028 (1009) & 1.28 (0.2)\\
MCI conv& 71 & 75.4 (6.7) & 40 & 21.3 (5.4) & 5882 (8644) & NA\\
AD & 98 & 74.7 (8) & 40 & 28.6 (8) & 5709 (1105) & 1.59 (0.1)\\
\end{tabular}\caption{Baseline clinical and sociodemographic information for the study cohort. NA: no observations available for the considered group. NL: normal individuals, MCI: mild cognitive impairment, AD: Alzheimer's patients. }\label{sociodem}
\end{table}

\begin{table}[b]
\centering
\begin{tabular}{ c c c c c c c c c }
Ventr & Hippo & Ent & Whole Brain & ADAS13 & FAQ & RAVLT & AV45 & FDG \\
\hline
\multicolumn{9}{c}{Training data}\\
\hline
%2.4 (1.9) & 2.4 (1.9) & 2.4 (1.9) & 2.4 (1.9) &3.2 (2.3) & 3.2 (2.3) & 3.2 (2.3) & 1.3 (1)  & 1.9 (1.9) \\
2.4 (0) & 2.4 (0) & 2.4 (0) & 2.4 (0) &3.2 (0) & 3.2 (0) & 3.2 (0) & 1.3 (0)  & 1.9 (0) \\
\hline
\multicolumn{9}{c}{Testing data}\\
\hline
1.8 (.1) & 1.8 (.1) & 1.8 (.1) & 1.8 (.1) & 2.4 (0) & 2.5 (0) & 2.4 (0) & 1.3 (65)  & 1.6 (31) \\
\end{tabular}\caption{Average follow-up years and percentage of individuals with missing data for each biomarker (in parenthesis).}\label{Nobs}
\end{table}

\begin{figure}[t]
\centering
\includegraphics[width=12cm]{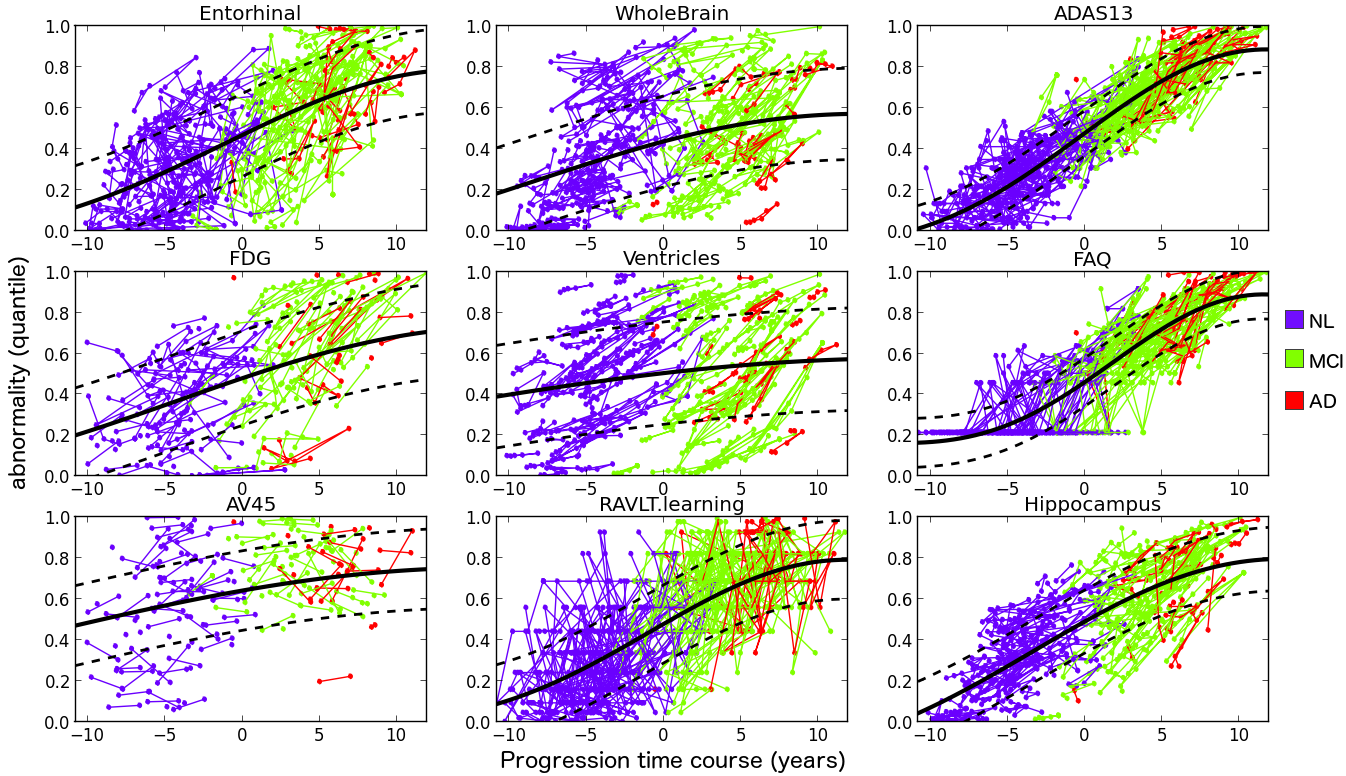}
\caption{Modelled biomarker progression in A$\beta$ amyloid positive individuals (solid/dashed lines: mean $\pm$ sd).  NL: normal individuals, MCI: mild cognitive impairment converted to AD, AD: Alzheimer's patients.}
\label{figModelAD}
\end{figure}

\begin{figure}[t]
\centering
\includegraphics[width=12cm]{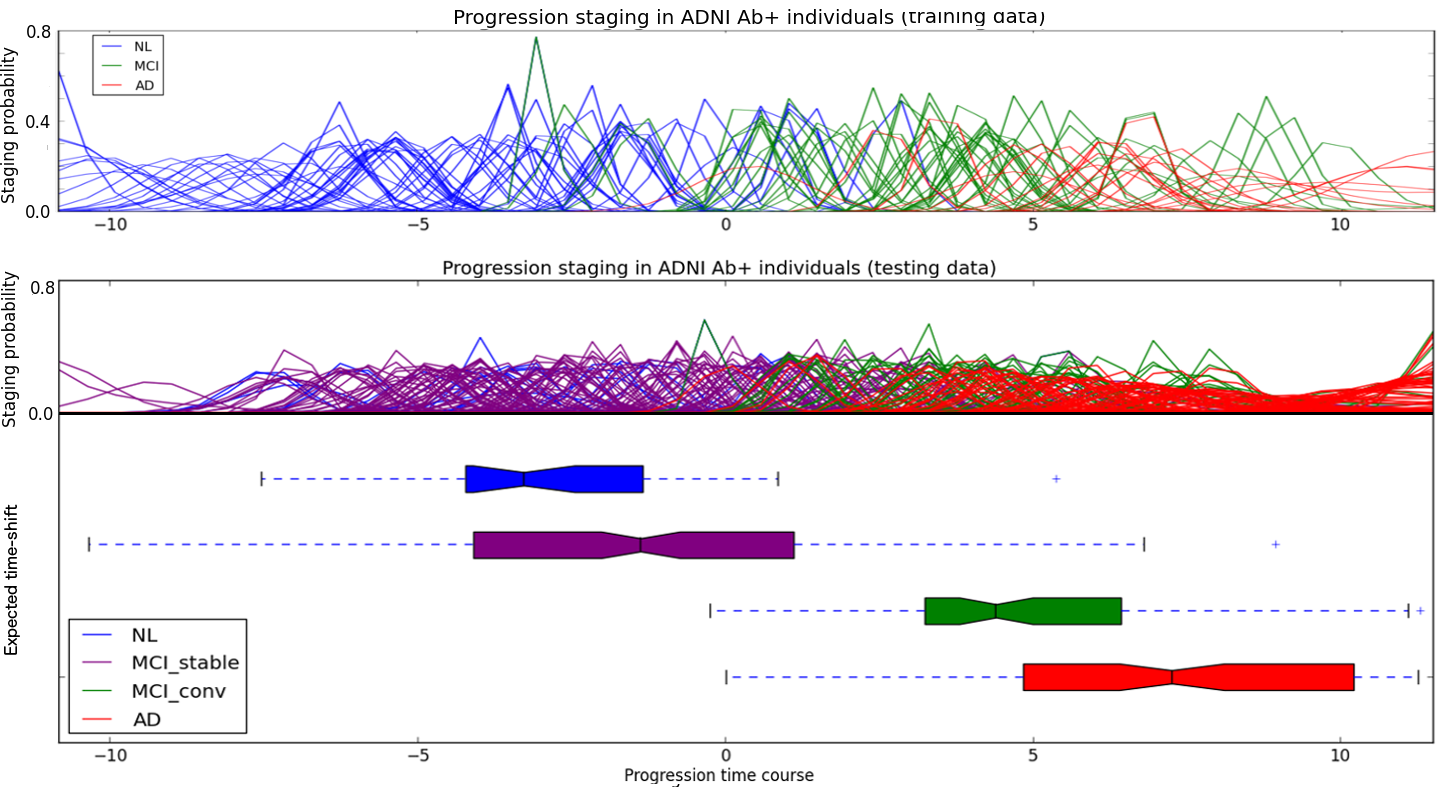}
\caption{Posterior prediction for the individual time shift in training (top) and testing (bottom) data. Healthy individuals are generally displaced at the early stages of the pathology, while the predictions for MCI and AD patients are associated with respectively intermediate and late progression stages.  NL: normal individuals, MCI: mild cognitive impairment, AD: Alzheimer's patients.  }
\label{figPrediction}
\end{figure}

The model was applied in order to estimate the temporal biomarker evolution and the disease stage associated with each individual in training  and testing set. The plausibility of the model was assessed by i) group-wise comparison of the predicted time-shift, ii) prediction of conversion to AD in the MCI testing group, and iii) correlation with respect to the time to AD diagnosis  for the MCI individuals subsequently converted to AD. We finally compared the progression modelled with our approach with respect to the one estimated  with the method proposed in \cite{donohue2014}. The method was applied to the training data by using the standard parameters defined in the R package \textsc{grace}\footnote{https://mdonohue.bitbucket.io/grace/}.\\
{\bf Results.}
The estimated biomarker progression (Figure \ref{figModelAD}) shows a biologically plausible description of the pathological evolution, compatible with previous findings in longitudinal studies in familial AD \cite{bateman2012}. The progression is defined on a time scale spanning roughly 20 years, and is characterized at the initial stages by high-levels of AV45, followed by an increase in ventricles volume and abnormality of FDG uptake. These  latter measures are however heterogeneously distributed across clinical groups, and with rather large variability. The evolution is further characterized by increasing abnormality of the volumetric measures, and finally by the steady worsening of neuropsychological scores such as FAQ.
Figure \ref{figPrediction} shows the posterior predicted distribution of the individual time shift.  Healthy individuals are associated with the early stages of the pathology in both training and testing data, while MCI and AD patients are characterized by respectively intermediate and late predicted progression stages. The group-wise comparison between the expected time-shifts is statistically significant between each group pairs ($p$ $<$1e-$4$). Furthermore, the temporal positioning of the non converting MCI lies between controls and MCI converters: when using the temporal cut-off based on the $10^{th}$ quantile of the time-shift distribution in the training AD population ($t=3.6$ years) we obtained an accuracy of $0.84$ for the discrimination between MCI converters and stable in the testing data. 
We also measured a negative correlation with respect to the time to AD diagnosis in training, testing, and pooled MCI converter groups, with $R^2$ respectively equal to $-0.20\, (p=0.1)$, $-0.28\, (p=0.01)$, and  $-0.28\, (p=0.001)$. 
Finally, when applying \cite{donohue2014} to the training data we measured a strong correlation between the resulting individual time-shifts and those obtained with our method ($R^2$ = $0.89$, $p$ $<$1e-$16$). Nevertheless, our estimates provided \emph{consistently larger effect sizes} for the group-wise separation: (1.96,1.36,0.57) with our methods and (1.74, 1.18, 0.47) with \textsc{grace} for AD vs NL, MCI vs NL, and AD vs MCI, respectively.

\section{Conclusions}
We proposed a unified GP-based approach to disease progression modeling from time-series of biomarker measurements enabling novel applications beyond the state-of-art, such as the probabilistic prediction of disease staging in unseen patients.
Furthermore, the model naturally accounts for missing data, and provides uncertainty quantification of the biomarker evolutions. The model provided remarkable modeling and predictive performance when tested on a large clinical cohort, and thus represents a promising instrument for the analysis of clinical trials data.

Similarly to \cite{donohue2014}, in this work we focused on the modeling of disease staging represented by a time translation. However, the proposed framework can naturally account for more complex time transformations, provided that a sufficient number of time points is available for each individual. %We also note that our approach is complementary to the spatio-temporal GP-based regression problem presented in \cite{lorenzi2015}.
Future extensions of this model will focus on the quantification of the effect of each biomarkers in the predictive performance, for example by integrating feature selection based on automatic relevance determination. Moreover the present work can be extended to model differential progressions underlying sub-pathologies, by framing the proposed random effect regression within the realm of Gaussian process mixture models. Finally, thanks to the flexibility of our framework, further extension of the model will enable to integrate  within (\ref{FixCovariance}) a spatio-temporal covariance model (such as the efficient Kronecker form of \cite{lorenzi2015}), to provide a unified framework for jointly modelling time series of images and scalar biomarkers data in a coherent fully Bayesian setting.

\section{Acknowledgments}
EPSRC grants EP/J020990/01 and EP/M020533/1 support DCA and SO's work on this topic. ML, DCA and SO also received support from the European Union's Horizon 2020 research and innovation programme under grant agreement No 666992 (EuroPOND) for this work. MF gratefully acknowledges support from the AXA Research Fund.
\bibliographystyle{splncs}
\bibliography{myrefs}

\begin{thebibliography}{10}
\providecommand{\url}[1]{\texttt{#1}}
\providecommand{\urlprefix}{URL }

\bibitem{bateman2012}
Bateman, R.J., Xiong, C., Benzinger, T.L., et~al.: Clinical and biomarker
  changes in dominantly inherited {A}lzheimer's disease. New England Journal of
  Medicine  367(9),  795--804 (2012)

\bibitem{bilgel2015}
Bilgel, M., Jedynak, B., Wong, D.F., Resnick, S.M., Prince, J.L.: Temporal
  trajectory and progression score estimation from voxelwise longitudinal
  imaging measures: Application to amyloid imaging. In: IPMI. pp. 424--436.
  Springer (2015)

\bibitem{donohue2014}
Donohue, M.C., Jacqmin-Gadda, H., Le~Goff, M., et~al.: Estimating long-term
  multivariate progression from short-term data. Alzheimer's \& Dementia
  10(5),  S400--S410 (2014)

\bibitem{fonteijn2011}
Fonteijn, H.M., Clarkson, M.J., Modat, M., et~al.: An event-based disease
  progression model and its application to familial {A}lzheimer's disease. In:
  IPMI. pp. 748--759. Springer (2011)

\bibitem{guerrero2016}
Guerrero, R., Schmidt-Richberg, A., Ledig, C., Tong, T., Wolz, R., Rueckert,
  D.: Instantiated mixed effects modeling of {A}lzheimer's disease markers.
  NeuroImage  142,  113--125 (2016)

\bibitem{kneip1988}
Kneip, A., Gasser, T.: Convergence and consistency results for self-modeling
  nonlinear regression. The Annals of Statistics pp. 82--112 (1988)

\bibitem{lorenzi2015}
Lorenzi, M., Ziegler, G., Alexander, D.C., Ourselin, S.: Efficient {G}aussian
  process-based modelling and prediction of image time series. In: IPMI. pp.
  626--637. Springer (2015)

\bibitem{minka2001expectation}
Minka, T.P.: Expectation propagation for approximate bayesian inference. In:
  Proceedings of the Seventeenth conference on Uncertainty in artificial
  intelligence. pp. 362--369. Morgan Kaufmann Publishers Inc. (2001)

\bibitem{rasmussen2006gaussian}
Rasmussen, C.E.: Gaussian processes for machine learning. Springer (2006)

\bibitem{riihimaki2010gaussian}
Riihim{\"a}ki, J., Vehtari, A.: Gaussian processes with monotonicity
  information. In: AISTATS. vol.~9, pp. 645--652 (2010)

\bibitem{schiratti2015}
Schiratti, J.B., Allassonniere, S., Routier, A., Colliot, O., Durrleman, S.: A
  mixed-effects model with time reparametrization for longitudinal univariate
  manifold-valued data. In: IPMI. pp. 564--575. Springer (2015)

\bibitem{yang2011}
Yang, E., Farnum, M., Lobanov, V., et~al.: Quantifying the pathophysiological
  timeline of {A}lzheimer's disease. Journal of Alzheimer's Disease  26(4),
  745--753 (2011)

\bibitem{younes2014}
Younes, L., Albert, M., Miller, M.I., Team, B.R., et~al.: Inferring changepoint
  times of medial temporal lobe morphometric change in preclinical
  {A}lzheimer's disease. NeuroImage: Clinical  5,  178--187 (2014)

\end{thebibliography}

\end{document}